\documentclass[12pt,english]{article}
\usepackage{
amsmath,
amssymb,
babel,
braket,
booktabs,
caption,
cite,
float,
graphbox,
graphicx,
hyperref,
pgfplots,
siunitx,
subfig,
subfloat,
tabu,
xcolor,
xparse,
ytableau
}

\hypersetup{colorlinks=true,
	linkcolor={red!30!black},
    	citecolor={blue!50!black},
    	urlcolor={blue!80!black}
   	 }

\newcommand{\eqb}{\begin{equation}}
\newcommand{\eqe}{\end{equation}}
\newcommand{\dmb}{\begin{displaymath}}
\newcommand{\dme}{\end{displaymath}}

\newcommand{\eab}{\begin{eqnarray}}
\newcommand{\eae}{\end{eqnarray}}

\newcommand{\be}{\begin{equation}}
\newcommand{\ee}{\end{equation}}
\newcommand{\sgn}{\text{sgn}\,}

\RenewDocumentCommand\[{}{\begin{equation}}
\RenewDocumentCommand\]{}{\end{equation}}
\NewDocumentCommand\der{}{\mathrm{d}}

\NewDocumentCommand\intkern{}{\int\kern-5pt}
\NewDocumentCommand\mat{m}{\left(\begin{matrix}#1\end{matrix}\right)}

\begin{document}
\begin{titlepage} 

\vspace{0.6cm}

\begin{center}
{\Large {Massive fluctuations in deconfining SU(2) Yang-Mills thermodynamics} \vspace{1.5cm}
}\\
{\Large{} }{\large {Ingolf Bischer} }
\par\end{center}
\vspace{1cm}
\begin{center}
Institut f\"ur Theoretische Physik, \\
Universit\"at Heidelberg, \\
Philosophenweg 16, 69120 Heidelberg, Germany
\end{center}

\vspace{1cm}
 
\begin{abstract}
We review how vertex constraints inherited from the thermal ground state strongly reduce the integration support of loop four-momenta associated with massive quasi-particles in bubble diagrams constituting corrections to the free thermal quasi-particle pressure. 
In spite of the observed increasingly suppressing effect when increasing 2-particle-irreducible (2PI) loop order, a quantitative analysis enables us to disprove the conjecture voiced in hep-th/0609033 that the loop expansion would terminate at a finite order.
This reveals the necessity to investigate exact expressions of (at least some) higher-loop order diagrams. Explicit calculation shows that although the behaviour of the 2PI three-loop contribution at low temperatures displays hierarchical suppression compared to lower loop-orders, its high-temperature expression instead dominates all lower orders. However, an all-loop-order resummation of a class of 2PI bubble diagrams is shown to yield an analytic continuation of the low-temperature hierarchy to all temperatures in the deconfining phase.
\end{abstract}
\end{titlepage}

\tableofcontents
\section{Introduction}
There is a variety of topologically non-trivial solutions to classical equations of motion in SU(2) gauge theory on a flat Euclidean spacetime manifold. That the trivial vacuum may not be the relevant one at non-zero temperature becomes apparent in the problems of the standard perturbative approach, in particular in the infrared problem already pointed out by A. D. Linde in 1980 \cite{Linde1980}. Divergences in the soft-magnetic sector, as encountered in small-coupling expansions at high temperature \cite{thermalPT1,thermalPT2,thermalPT3,thermalPT4,thermalPT5,thermalPT6,thermalPT7}, motivated by asymptotic freedom \cite{gross-wilczek,politzer,khriplovich}, invalidate the perturbative expansion starting at some finite order \cite{thermalPT8} and hint at relevant substructures that are missed. Indeed, lattice-based studies relate topological configurations to fundamental properties of Yang-Mills theory \cite{reinhardt,faber,muller}.
An approach to finding a thermal ground state estimate that includes gauge field configurations of non-trivial topology reveals that Harrington-Shepard (anti)calorons \cite{HS1977} of topological charge $|k|=1$ are the constituents of this ground state with spatially densely packed centers and overlapping peripheries. Their contribution is manifest in the non-triviality of the spatial and scale-parameter average (spatial coarse-graining)  of the two-point field-strength correlator in association with the magnetic field of an (anti)caloron \cite{hofmannbook}. Lattice gauge theory qualitatively reproduces certain aspects 
of this correlation in infrared sensitive thermodynamical quantities such as the pressure, provided the \textsl{differential} method is used which appeals to the non-perturbative beta function \cite{Deng1988,Brown1988}. However, this function needs to be approximated. On the other hand, the \textsl{integral} method \cite{engels}, which does not rely on the beta function but introduces an integration constant, yields results that are largely disparate, the reason being the choice of integration constant (no negative pressure) and finite-volume artifacts \cite{hofmann05}.

In this work, we give an overview of recent proceedings in the treatment of radiative corrections to the pressure of this thermal ground state beyond two-loop order. These corrections are obtained by a loop expansion of the three effective gauge fields (quasi-particles) obtained after coarse-graining over the ground state constituent configurations, two of which become massive by an adjoint Higgs mechanism. We find that resummation of infinitely many diagrams is necessary to obtain a finite result which after resummation is well-controlled in the case of the diagrams treated here. A much more detailed and technical presentation of our results can be found in \cite{BGH}.

This work is structured as follows. In \autoref{sec:sign-constraints}, we present a non-exhaustive way of using constraints in the massive sector to reduce the number of possible loop-momentum configurations in bubble diagrams in a purely combinatorical way. In \autoref{sec:upto3l}, we state the contributions of all bubble diagrams in the massive sector up to three-loop and conclude why resummation is necessary. This resummation of a particular family of diagrams is finally demonstrated in \autoref{sec:3} and followed by a summary and conclusions in \autoref{sec:4}

\section{Sign constraints in massive bubble diagrams}\label{sec:sign-constraints}
In this section, we explain the origin and structure of sign constraints on massive quasi-particle loop momenta mediated by four-vertices. We state the results of an efficient book-keeping explained in \cite{BGH} in terms of the ratio of the number of non-excluded sign configurations and the number of a priori possible sign configurations. To close the section, an explanation of why non-vanishing diagrams exist at any finite loop order is given.

\begin{figure}[H]
\centering{
{\Large $\frac{1}{48}\cdot\,$}
\includegraphics[align=c]{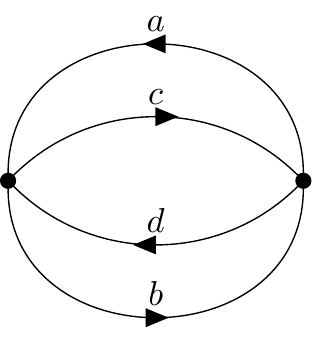}
\phantom{{\Large $\frac{1}{48}\cdot\,$}}
\caption{The only 2PI three-loop diagram (symmetry factor $1/48$, $R=1/6$)}
\label{fig:3-loop-1}
}
\end{figure}

The full set of Feynman rules for the quasi-particles populating the thermal ground state in the deconfining phase is listed in \cite{hofmannbook}. Here, we restrict the discussion to 2PI diagrams, by which we mean bubble diagrams that do not become 1PI contributions to a polarisation tensor upon cutting any single line, including only the two massive fields (corresponding to two su(2) algebra directions that are broken by the thermal ground state and obtain a mass by an adjoint Higgs mechanism). This implies that only four-vertices may appear.
The first important fact for what follows is that those massive fields propagate strictly on-shell
\[
p^2=m^2=4e^2|\phi|^2,
\]
where $p$ is any four-momentum, $m$ is the mass, $e$ is the effective gauge coupling, and $|\phi|$ is the gauge invariant modulus of the inert, adjoint scalar field associated with densely packed (anti)caloron centers in the thermal ground state 
\cite{hofmannbook,Entropy2016} which sets the scale of maximal resolution.
The second important fact is that the scattering channels at four-vertices are restricted not to resolve higher energies than this scale. By this we mean that each four-vertex hosts a superposition of channels corresponding to the three Mandelstam variables $s$, $t$, and $u$ constrained by $|s|,|t|,|u|\leq|\phi|^2$. By virtue of the on-shellness, each constraint on a Mandelstam variable implies a restriction of the energy-signs of the respective loop momenta according to \cite{hofmann-krasowski}
\begin{align}\label{eq:generalconstraints1}
|(p+q)^2|\leq|\phi|^2 \quad &\Rightarrow \quad \sgn(p_0)=-\sgn(q_0)\,,\\
\label{eq:generalconstraints2}
|(p-q)^2|\leq|\phi|^2 \quad &\Rightarrow \quad \sgn(p_0)=\sgn(q_0)\,.
\end{align}
Hence, for all scattering channel combinations in a diagram, one can exclude several sign configurations. We define the ratio $R$ of a diagram by the sum (over channel combinations) of the numbers of non-excluded sign configurations divided by the number of a priori possible sign configurations times the number of channel combinations ($2^{2n}\cdot3^n$, where $n$ denotes the number of vertices)). In \autoref{fig:3-loop-1} through \autoref{fig:6-loop-2-4} we give all 2PI diagrams up to six-loop order and their respective values of $R$.  All results are listed in \autoref{tab:signconstraints}.
\begin{figure}[H]
\centering{
{\Large $\frac{1}{48}\cdot$}
\includegraphics[align=c]{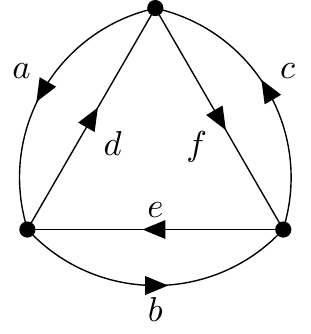}
\phantom{{\Large $\frac{1}{48}\cdot$}}
\caption{The only 2PI four-loop diagram (symmetry factor $1/48$, $R=\frac{5}{108}=0.0463$)}
\label{fig:4-loop-1}
}
\end{figure}
\begin{figure}[H]
{\centering{{\Large$\frac{1}{128}\cdot$}
\includegraphics[scale=1,align=c]{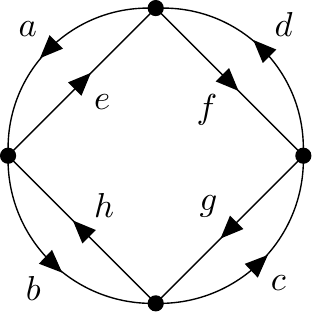}
\phantom{{\Large$\frac{1}{128}\cdot$}}
\qquad
{\Large$\frac{1}{32}\cdot$}
\includegraphics[scale=1,align=c]{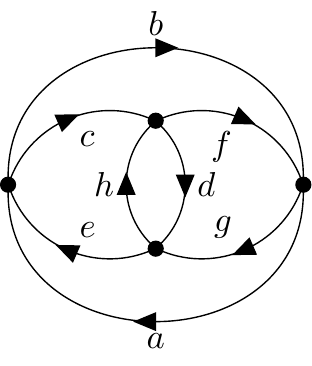}
\phantom{{\Large$\frac{1}{128}\cdot$}}
\caption{The first and second 2PI five-loop diagram (symmetry factors $1/128$ (first), $1/32$ (second) and $R=1/72=0.0139$ (first), $R=1/81=0.0123$ (second)).}
\label{fig:5-loop}}}
\end{figure}
\begin{figure}[H]
{\centering{{\Large$\frac{1}{320}\cdot$}
\includegraphics[scale=1,align=c]{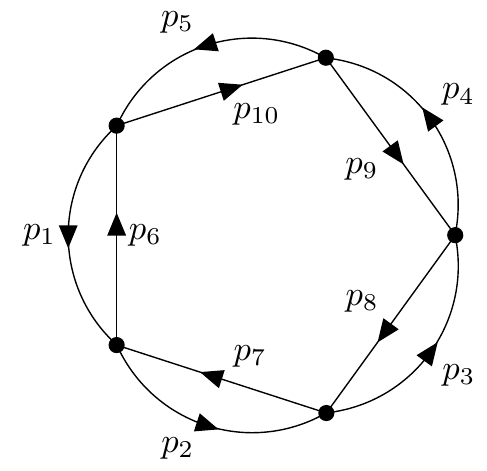}
\phantom{{\Large$\frac{1}{320}\cdot$}}
\caption{The first and most symmetric 2PI six-loop diagram (symmetry factor $1/320$, $R=17/3888=0.0044$).}
\label{fig:6-loop-1}}}
\end{figure}
\begin{subfigures}
\begin{figure}[H]
{\centering{
{\Large$\frac{1}{32}\cdot$}
\includegraphics[align=c]{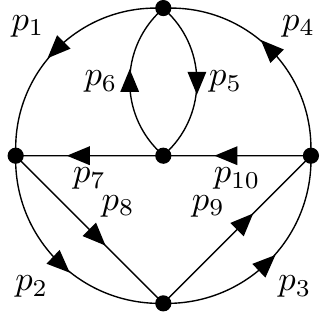}
\phantom{{\Large$\frac{1}{32}\cdot$}}
\qquad
{\Large$\frac{1}{16}\cdot$}
\includegraphics[align=c]{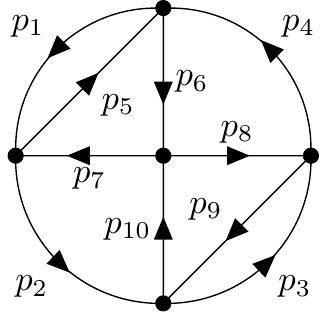}
\phantom{{\Large$\frac{1}{16}\cdot$}}
\caption{The second and third 2PI six-loop diagrams (symmetry factors $1/32$ and $1/16$, $R=7/1944=0.0036$ and $R=13/3888=0.0033$).}
\label{fig:6-loop-2}}}
\end{figure}
\begin{figure}[H]
{\centering{
{\Large$\frac{1}{120}\cdot$}
\includegraphics[align=c]{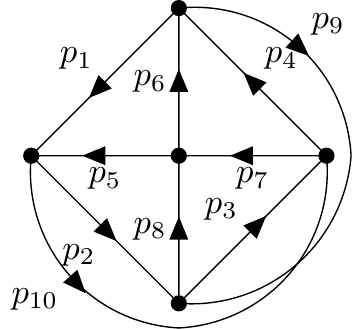}
\phantom{{\Large$\frac{1}{120}\cdot$}}
\caption{The fourth 2PI six-loop diagram (symmetry factor $1/120$, $R=1/324=0.0031$). This is the only non-planar diagram up 
to six-loop order.}
\label{fig:6-loop-3}}}
\end{figure}
\label{fig:6-loop-2-4}
\end{subfigures}

\begin{table}
\begin{tabu}{rrX[r]X[r,$]X[r, $]}
\toprule
Loop number & Diagram number & $R$ & S^{-1} & R\cdot S\\
\midrule
3	&1	&0.1667 & 48 & 0.00347222\\\addlinespace[2pt]
4	&1	&0.0463 & 48 & 0.00096451\\\addlinespace[2pt]
5	&1	&0.0139 & 128 & 0.00010851\\
5	&2	&0.0123 & 32 & 0.00038580\\\addlinespace[2pt]
6	&1	&0.0044 & 320 & 0.00001366\\
6	&2	&0.0036 & 32 & 0.00011253\\
6	&3	&0.0033 & 16 & 0.00020898\\
6	&4	&0.0033 & 120 & 0.00002572\\
\bottomrule
\end{tabu}
\caption{Ratio $R$ of allowed vs. a priori possible energy-sign and scattering-channel combinations 
for 2PI bubble diagrams up to six loops. $S$ denotes a diagram's symmetry factor.}
\label{tab:signconstraints}
\end{table}

In agreement with a simple counting argument given in \cite{hofmann06}, we observe a monotonic decrease of $R$ with increasing loop order. However, none of the diagrams become completely excluded. Indeed, one can show that diagrams with non-excluded sign configurations (i.e. diagrams with $R>0$) exist at any finite loop order \cite{BGH}. This is most transparent in the class of diagrams of highest symmetry, namely \autoref{fig:3-loop-1} and the diagrams symmetric under the $n$-th dihedral group, \autoref{fig:4-loop-1}, \autoref{fig:5-loop} (left side), and \autoref{fig:6-loop-1}. In this class, there is a vertex channel combination such that only the two momenta connecting the same two vertices appear as pairs in a constraint, e.g. for the diagram in \autoref{fig:6-loop-1} the configuration
\[
|(p_1-p_6)^2|=|(p_2-p_7)^2|=|(p_3-p_8)^2|=|(p_4-p_9)^2|=|(p_5-p_{10})^2|\leq|\phi|^2\,,
\]
where the equalities stem from momentum conservation at each vertex. One independent constraint, however, is not sufficient to exclude all sign configurations and it follows that $R>0$. In the cases of lower-symmetry diagrams in \autoref{fig:5-loop} (right side) and \autoref{fig:6-loop-2-4} there are fewer non-excluded configurations compared to \autoref{fig:5-loop} (left side) and \autoref{fig:6-loop-1} respectively, so indeed symmetry appears to be associated with the ratio $R$.

Despite this drawback, the actual order of magnitude of the higher loop order diagrams is not at all obvious from these sign considerations. Thus it is necessary to consider full expressions of the loop integrals to make definite statements about the convergence properties of the loop expansion. In the next section, we hence discuss the results of explicit calculations up to three-loop order which display hierarchical ordering at low temperatures but a dominating three-loop contribution at high temperatures.

\section{The massive sector up to three loops} \label{sec:upto3l}

\subsection{One-loop pressure}

In general, the expansion of the deconfining pressure in SU(2) Yang-Mills thermodynamics reads
\[
P = P_\mathrm{gs} + P_{\mathrm{1-loop}} + \Delta P\,,
\]
where $P_\mathrm{gs}=-4\pi\Lambda^3T$ denotes the negative contribution from the ground-state estimate, $P_\mathrm{1-loop}$ represents the pressure exerted by non-interacting thermal quasiparticles (one-loop), and $\Delta P$ summarises all radiative corrections as expanded in ascending loop orders. Here, $\Lambda$ denotes the Yang-Mills scale. 
Unlike in standard perturbation theory, the radiative corrections do not represent an asymptotic (power) series in the coupling constant. As hinted in \autoref{sec:sign-constraints}, the usefulness of loop ordering in this case stems from the increasing number of constraints on loop integrations with increasing loop order. Loosely speaking, the quantity which serves as a (non-local) expansion parameter is the highly constrained volume of loop momenta over the unconstrained volume. The expectation consistent with previous calculations \cite{herbst-hofmann-rohrer,SGH2007} is that fixed-order contributions to $\Delta P$ decrease strongly enough with increasing loop order and number of constraints as to render the expansion convergent in the standard mathematical sense. As we discuss below, however, this is not the case at high temperatures, where resummation techniques have to be applied in order to extend the convergent low-temperature behaviour.
On the level of free quasiparticles, the trace anomaly of the energy-momentum tensor, which rises linearly in $T$, is invoked by both $P_\mathrm{gs}$ and the massive contribution of $P_\mathrm{1-loop}$ \cite{giacosa-trace}.

Restricting ourselves to the massive sector only, the one-loop pressure reads \cite{hofmannbook}
\begin{align}
\label{eq:oneloop-pressuremassive}
P(\lambda)|_\mathrm{1-loop} &= -\Lambda^4\frac{12\lambda^4}{(2\pi)^6}\bar P(2a)\,, 
\end{align}
where 
\begin{align}\label{eq:dimpress}
\bar P(y) &=\int_0^\infty \der x\, x^2\log\left[1 - e^{-\sqrt{x^2+y^2}}\right]\,,
\end{align} 
$\lambda\equiv \frac{2\pi T}{\Lambda},$ and $a\equiv\frac{m}{2T}$. The one-loop pressure rapidly saturates into the $T^4$ behaviour of the Stefan-Boltzmann limit. Notice that even at high temperatures, where this limit is approached in a power-like way, the number of independent polarisations is six rather than four due to the thermal ground state minutely breaking the original gauge symmetry. This means that including the massless gauge mode one arrives at eight rather than six polarisations as generally utilised in perturbative and phenomenological ``bag model'' \cite{satz} calculations, the two additional degrees of freedom originating from the scalar magnetic monopole and its antimonopole \cite{hofmannbook}. The thermal ground-state contribution $P_\mathrm{gs}$ would be modelled by a temperature-dependent bag pressure.

\subsection{Two-loop correction}
\begin{figure}
\centering{
{\Large $\frac{1}{8}\cdot\,$}
\includegraphics[align=c]{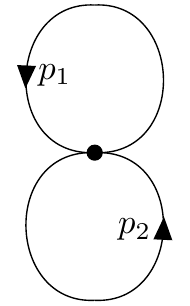}
\phantom{{\Large $\frac{1}{8}\cdot\,$}}
\caption{The two-loop diagram for the pressure in the massive sector of deconfining SU(2) Yang-Mills thermodynamics (symmetry factor $1/8$).}
\label{fig:2-loop-2}
}
\end{figure}
The pressure contribution associated with the two-loop diagram in \autoref{fig:2-loop-2} reads \cite{herbst-hofmann-rohrer,SGH2007}
\begin{multline}\label{eq:2-loop-exact}
\Delta P|_\mathrm{2-loop} = \frac{-2e^2T^4}{\lambda^6}\intkern\der r_1\der r_2\der\cos\theta\frac{r_1^2r_2^2}{\sqrt{r_1^2+m^2}\sqrt{r_2^2+m^2}}\\
\times\left[ 14-2\frac{k^4}{m^4}\right]
 n_B\left(2\pi\sqrt{\frac{r_1^2+m^2}{\lambda^3}}\right)n_B\left(2\pi\sqrt{\frac{r_2^2+m^2}{\lambda^3}}\right) \,,
\end{multline}
where 
\[
k^2\equiv p_1p_2 = -\sqrt{r_1^2+m^2}\sqrt{r_2^2+m^2} -r_1r_2\cos\theta
\] 
is defined as the Lorentz-invariant product of the dimensionless\footnote{We normalise physical four-momentum components $P^\mu$ by $|\phi|$ to arrive at dimensionless components $p^\mu$. Likewise, the physical mass is made dimensionless: $m=2e$.}
loop four-momenta $p_1$ and $p_2$, $r_1=|\mathbf{p}_1|$ and $r_2=|\mathbf{p}_2|$ denote the moduli of their spatial parts, $n_B(x) = (\exp(x)-1)^{-1}$ refers to the Bose-Einstein distribution function, and 
the integration is subject to the constraint
\[
|2m^2-2\sqrt{r_1^2+m^2}\sqrt{r_2^2+m^2}-2r_1r_2\cos\theta|\leq1\,.
\]
In \autoref{fig:results} on the left hand side, the temperature dependence of the numerical integrations in \eqref{eq:2-loop-exact} and \eqref{eq:oneloop-pressuremassive} is shown in terms of their ratio.

\subsection{Three-loop correction}
The pressure contribution associated with the diagram in \autoref{fig:3-loop-1} has been calculated in \cite{BGH}. After relabelling ($a,b,c,d\rightarrow p_1,p_3,p_2,p_4$) and in terms of dimensionless momenta it reads
\begin{multline}\label{eq:3-loop-start}
\Delta P|_{\mathrm{3-loop}} = i\frac{\Lambda^4}{48\lambda^2}e^4\frac1{(2\pi)^6}\sum_{\mathrm{signs}}\intkern\der\theta_1\der\varphi_1\der r_1\der r_2\der\theta_3\sum_{\{r_3\}}r^2_1r^2_2r^2_3\sin\theta_1\sin\theta_3\\
\times P(p_i)\frac{n_B'(r_1)n_B'(r_2)n_B'(r_3)n_B'(r_4)}{8|p^0_1p^0_2p^0_3p^0_4|}\,,
\end{multline}
The first sum in \eqref{eq:3-loop-start} runs over allowed sign combinations for $p^0_i$, $i=1,\dots,4$.
All four-momenta $p_i\equiv (p^0_i, \mathbf{p}_i)$ are on-shell, $|p_i^0|\equiv\sqrt{\mathbf{p}_i^2+m^2}$, and are parametrised as 
\begin{equation*}
p_4\equiv p_2+p_3-p_1,\quad \mathbf{p}_2\equiv\mat{0\\0\\r_2},\quad \mathbf{p}_3\equiv r_3\mat{0\\\sin\theta_3 \\\cos\theta_3}, \quad \mathbf{p}_1\equiv r_1\mat{\sin\theta_1\cos\varphi_1\\ \sin\theta_1\sin\varphi_1 \\ \cos\theta_1}\,.
\end{equation*}
In the equivalent cases $ss$, $tt$, $uu$ (diagonal), the integration is constrained by
\[\label{eq:constraint-ss}
|(p_1+p_4)^2|=|(p_2+p_3)|^2\leq 1\phantom{.}.
\]
Summing over these cases, the resulting contribution to $\Delta P|_{\mathrm{3-loop}}$ is denoted by $\linebreak[0]1/3\Delta P|_{\mathrm{3-loop},ss}$.
On the other hand, for the equivalent cases $st$, $su$, $tu$, $ts$, $us$, $ut$ (off-diagonal) the constraints on the integration read
\[\begin{split}\label{eq:constraint-st}
|(p_1+p_4)^2|&=|(p_2+p_3)|^2\leq 1\,,\\
|(p_1-p_2)^2|&=|(p_3-p_4)|^2\leq 1\,.
\end{split}\]
The sum of these cases amounts to $2/3\Delta P|_{\mathrm{3-loop},st}\,$, such that
\[\label{eq:pressure-3-loop}
\Delta P|_\mathrm{3-loop} = \frac13\Delta P|_{\mathrm{3-loop},ss} + \frac23\Delta P|_{\mathrm{3-loop},st}\,.
\]
The second sum in \eqref{eq:3-loop-start} runs over all solutions in $r_3$ of the equation
\begin{multline}\label{eq:r3}
\sgn(p^0_2)\sqrt{r_2^2+m^2}+\sgn(p^0_3)\sqrt{r^2_3+m^2}-\sqrt{r_1^2+m^2}=
-\left[r^2_1+r^2_2+r^2_3 \right.\\\left. - 2 r_1r_2\cos\theta_1  -2r_1r_3(\sin\varphi_1\sin\theta_1\sin\theta_3+\cos\theta_1\cos\theta_3)+2r_2r_3\cos\theta_3 + m^2\right]^{1/2}\,.
\end{multline}
The polynomial $P(\{p_i\})$ reads
\begin{multline}\label{eq:polynomial}
P(\{p_i\})= 144 -12\frac1{m^4}\left\{\vphantom{\frac12}(p_1p_2)^2+(p_1p_3)^2+(p_1p_4)^2+(p_2p_3)^2+(p_2p_4)^2+(p_3p_4)^2 \right\} \\
+36\frac1{m^6}\left\{\vphantom{\frac12}(p_1p_2)(p_1p_3)(p_2p_3)+(p_1p_2)(p_1p_4)(p_2p_4)
\right.\\
\left.\vphantom{\frac12}
+(p_1p_3)(p_1p_4)(p_3p_4)+(p_2p_3)(p_2p_4)(p_3p_4)\right\}\\
+12\frac1{m^8}\left\{\vphantom{\frac12}(p_1p_2)^2(p_3p_4)^2+(p_1p_3)^2(p_2p_4)^2+(p_1p_4)^2(p_2p_3)^2 - (p_1p_2)(p_1p_3)(p_2p_4)(p_3p_4)
\right.\\
\left.\vphantom{\frac12} -(p_1p_2)(p_1p_4)(p_2p_3)(p_3p_4) - (p_1p_3)(p_1p_4)(p_2p_3)(p_2p_4)\right\},
\end{multline}
and the Bose-Einstein distribution shorthand notation is
\[
n_B'(r)\equiv n_B\left(2\pi\sqrt{r^2+m^2}/\lambda^{3/2}\right).
\]
This complicated expression can be evaluated by Monte Carlo methods for low temperatures (close to the critical temperature $\lambda_c=13.87$) due to the Bose suppression of large spatial momenta $r_1$ and $r_2$. However, the high-temperature limit is inaccessible in this way, since the maxima of the product of the Bose functions $n_B'(r_i)$ and polynomials in $r_i$ get shifted to large $r_i$ like $\lambda^{3/2}$. Analysing the properties of the constraints, it is possible to obtain analytic high temperature expressions for both diagonal and off-diagonal contribution whose leading powers in $\lambda$ read \cite{BGH}
\[
\begin{split}\label{eq:approxintss3}
\frac{1}{3}\Delta P|_\mathrm{\mathrm{3-loop},ss} &\approx i\Lambda^4\frac{1}{3375}\frac{1}{(2\pi)^{15}}\frac{1}{m^4}\left(1+\frac{1}{4m^2}\right)\left(\pi^4-90\zeta(5)\right)^2\lambda^{13}\\ 
&\equiv ic_{13}\Lambda^4\lambda^{13}\,,
\end{split}
\]
where $\zeta(x)$ denotes Riemann's zeta function and the numerical value of the coefficient is $c_{13}=5.2968\cdot10^{-20}$ and
\[
\begin{split}\label{eq:approxintst3}
\frac{2}{3}\Delta P|_{\mathrm{3-loop},st} &\approx  i\Lambda^4e^4\frac{C}{(2\pi)^4}\frac{1}{12m^4}\lambda^4\approx i\Lambda^4 \lambda^4\cdot2.2011\cdot 10^{-12}\,,
\end{split}
\]
where
\[
\begin{split}
C&\equiv\frac{1}{2304}\frac{1}{(2\pi)^5}\sqrt{\frac4{m^2}+\frac1{m^4}}\left(132+\frac{72}{64}\frac1{m^4}+\frac{3}{1024}\frac{1}{m^8}\right)\,.
\end{split}
\]
The numerical values are obtained using the high-temperature plateau value of the mass and coupling $m=2e=2\sqrt{8}\pi$.

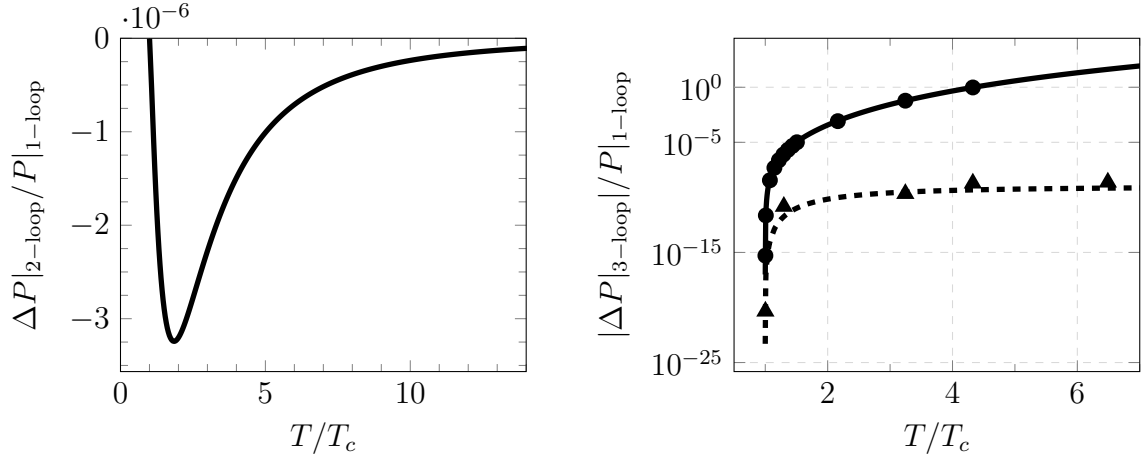
\begin{figure}
\centering{
    \begin{tikzpicture}
      \begin{axis}[
          width=0.45\linewidth, 
          xlabel=$T/T_c$, 
          ylabel=$\Delta P|_{\mathrm{2-loop}}/P|_{\mathrm{1-loop}}$,
          minor x tick num={4},
          minor y tick num={3},
          xmin=0, xmax=14,
         ymax=0,
        ]
        \addplot[
        smooth,
        line width=2pt,
        ] 
        table{graphs/2-loop-plot.dat}; 
      \end{axis}
\end{tikzpicture}
    \hphantom{$-1$}
\begin{tikzpicture}
      \begin{semilogyaxis}[
          width=.45\linewidth, 
          grid=major, 
          grid style={dashed,gray!30}, 
          xlabel=$T/T_c$, 
          ylabel=$|\Delta P|_\mathrm{3-loop}|/P|_{\mathrm{1-loop}}$,
          minor x tick num={1},
          minor y tick num={4},
          extra y ticks={1},
          y label style={at={(0,0.5)},anchor=south},
          xmin=0.5, xmax=7,
        ]
        \addplot[
        line width=2pt,
        ] 
        table{graphs/3-loop-ss-results-approx.dat}; 
        
         \addplot[
         only marks,
        line width=2pt,
        ] 
        table{graphs/3-loop-ss-results-mc.dat}; 

        \addplot[
        smooth,
        dashed,
        line width=2pt,
        ] 
        table{graphs/3-loop-st-results-approx.dat}; 
        
        \addplot[
         only marks,
        line width=2pt,
         mark=triangle,
        ] 
        table{graphs/3-loop-st-results-mc.dat}; 
      \end{semilogyaxis}
    \end{tikzpicture}
    \hphantom{$10^{-25}$}
\caption{The two-loop pressure contribution (left) and moduli of the three-loop pressure corrections (right), $1/3\Delta P|_{\mathrm{3-loop},ss}$ (solid) and 
$2/3\Delta P|_{\mathrm{3-loop},st}$ (dashed), divided by the massive sector one-loop pressure $P|_{\mathrm{1-loop}}$. On the right, the continuous curves represent the analytical high-$T$ expressions from \cite{BGH}, while the dots and triangles are the respective $ss$ and $st$ Monte Carlo (MC) results.}
\label{fig:results}}
\end{figure}

In \autoref{fig:results}, we compare the results of the two-loop and three-loop expressions $\Delta P|_{\mathrm{2-loop}}$ and $\Delta P|_{\mathrm{3-loop}}=1/3\Delta P|_{\mathrm{3-loop},ss}+2/3\Delta P|_{\mathrm{3-loop},st}$ divided by the one-loop expression $P|_{\mathrm{1-loop}}$. We emphasise the excellent matching of the Monte Carlo results at low temperatures with the high-temperature approximations, displaying a consistent transition into the power laws \eqref{eq:approxintss3} and \eqref{eq:approxintst3}.
Firstly we note that in the 3-loop case the off-diagonal contribution $2/3\Delta P|_{\mathrm{3-loop},st}$ is subleading to the diagonal contribution $1/3\Delta P|_{\mathrm{3-loop},ss}$. This allows us to neglect the former in the following discussions, while we stress that the power of an additional independent vertex constraint is impressively demonstrated by a reduction of the power law from $\lambda^{13}$ to $\lambda^4$.

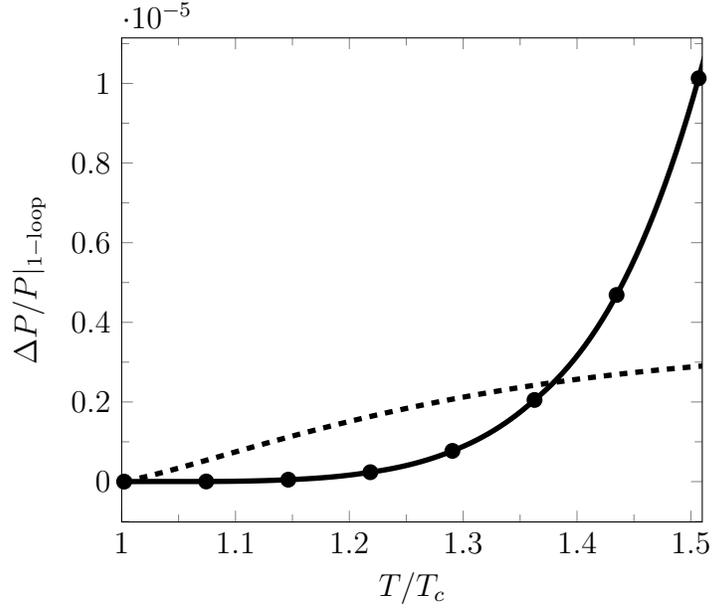
\begin{figure}
{\centering
    \begin{tikzpicture}
      \begin{axis}[
          width=.6\linewidth, 
          xlabel=$T/T_c$, 
          ylabel=$\Delta P/P|_{\mathrm{1-loop}}$,
          minor x tick num={1},
          minor y tick num={1},
          xmin=1, xmax=1.51,
        ]
        \addplot[
        dashed,
        smooth,
        line width=2pt,
        ] 
        table{graphs/2l-transition.dat}; 
        
         \addplot[
         only marks,
        line width=2pt,
        ] 
        table{graphs/3l-transition.dat}; 

        \addplot[
        smooth,
        line width=2pt,
        ] 
        table{graphs/3l-transition-interpolation.dat}; 
      \end{axis}
    \end{tikzpicture}
    \hphantom{$0.6$}
\caption{Monte Carlo results of $\Delta P|_{\mathrm{3-loop}}/P|_{\mathrm{1-loop}}$ close to $\lambda_c$ (dots). The solid line is a smooth interpolation of the latter while the dashed line represents $|\Delta P|_{\mathrm{2-loop}}|/P|_{\mathrm{1-loop}}$. 
}
\label{fig:3l-vs-2l}
}
\end{figure}

Comparing $|\Delta P|_{\mathrm{2-loop}}|$ and $|\Delta P|_{\mathrm{3-loop}}|$ with $P|_\mathrm{1-loop}\,$, apparently the high-tem\-perature behaviour of $|\Delta P|_{\mathrm{3-loop}}|$ is dramatically exceeding the lower orders. However, as shown in \autoref{fig:3l-vs-2l}, at low temperatures a hierarchical ordering $|\Delta P|_{\mathrm{3-loop}}|\ll |\Delta P|_{\mathrm{2-loop}}|$ is in fact observed. This leads us to the following conclusion: A fixed-order loop expansion is inappropriate at high temperatures. Instead, one needs to consider a resummation of diagrams with large contributions like the three-loop diagram which should then analytically continue the controlled low-temperature situation. This is demonstrated in the next section and amounts to the resummation of the family of dihedrally symmetric diagrams introduced in \autoref{sec:sign-constraints}. We will comment on the imaginary nature of some contributions after this resummation procedure.

\section{Resummation of the highest-symmetry diagrams}\label{sec:3}
In order to make sense of the high-temperature behaviour of the three-loop diagram, we consider a truncated version of the Dyson-Schwinger (DS) equation of the four-vertex which reads
\[
\label{eq:dsphi4}
\includegraphics[scale=0.6,align=c]{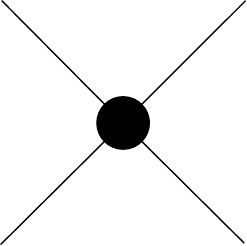} 
= \includegraphics[scale=0.6,align=c]{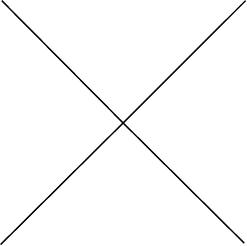} 
+ \includegraphics[scale=0.6,align=c]{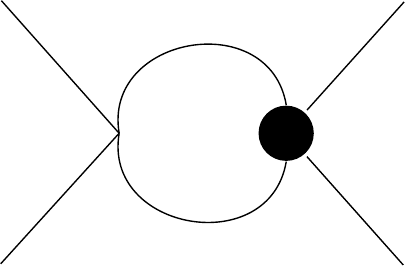},
\]
where undotted vertices are tree-level vertices, dotted vertices are (fully) resummed vertices and loop lines correspond to (fully) resummed propagators. For a non-vanishing result of the tree-level vertex (in the absence of massless fields), it is required that two external lines carry an algebra index of the first broken direction and the other two lines carry an index of the second broken direction. We asumme in the following that this tensorial structure also holds for the resummed vertex. This amounts to a scalar form factor $f(\lambda,i)$, $i=s,t,u$ multiplying the tree-level expression. Resummation of the propagators amounts to only mild deviations from the tree-level expressions \cite{hofmann06}. This justifies using the latter for further argumentation. Then \eqref{eq:dsphi4} has the interpretation of iteratively summing the following infinite number of diagrams
\[\label{eq:iteration}
\begin{split}
\includegraphics[scale=0.5,align=c]{feynman-graphs/DS-4-vertex-1}_{\,1}& 
= \includegraphics[scale=0.5,align=c]{feynman-graphs/DS-4-vertex-2}  + \includegraphics[scale=0.5,align=c]{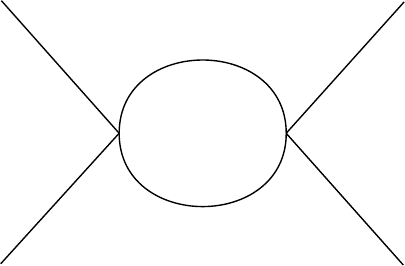} \\
\includegraphics[scale=0.5,align=c]{feynman-graphs/DS-4-vertex-1}_{\,2}&
=\includegraphics[scale=0.5,align=c]{feynman-graphs/DS-4-vertex-2}  + \includegraphics[scale=0.5,align=c]{feynman-graphs/DS-4-vertex-3}_{\,1}
=\includegraphics[scale=0.5,align=c]{feynman-graphs/DS-4-vertex-2}  + \includegraphics[scale=0.5,align=c]{feynman-graphs/DS-preiteration-1l} +\includegraphics[scale=0.5,align=c]{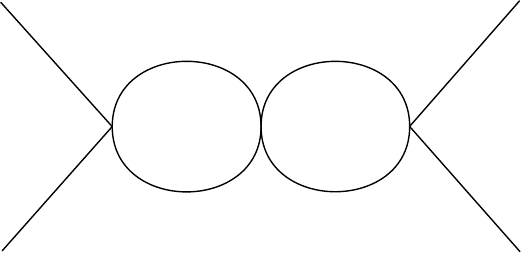}  \\
\includegraphics[scale=0.5,align=c]{feynman-graphs/DS-4-vertex-1}_{\,3}& 
=\includegraphics[scale=0.5,align=c]{feynman-graphs/DS-4-vertex-2}  + \includegraphics[scale=0.5,align=c]{feynman-graphs/DS-preiteration-1l} +\includegraphics[scale=0.5,align=c]{feynman-graphs/DS-preiteration-2l}+\includegraphics[scale=0.5,align=c]{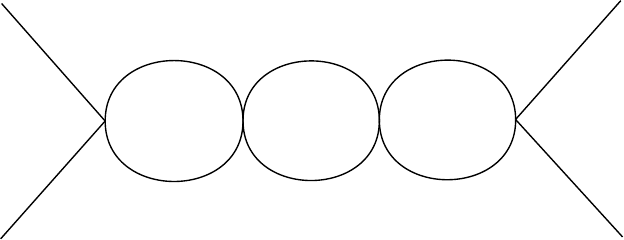}\\
&\dots
\end{split}
\]
When closing legs into two (extra) loops, this becomes the resummation of the class of dihedrally symmetric bubble diagrams:
\[
\includegraphics[scale=0.6,align=c]{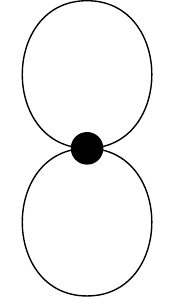} =
\includegraphics[scale=0.6,align=c]{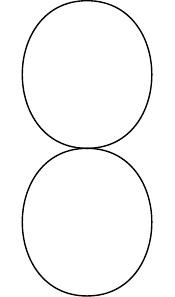}+
\includegraphics[scale=0.6,align=c]{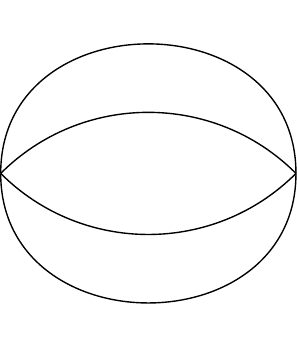}+
\includegraphics[scale=0.6,align=c]{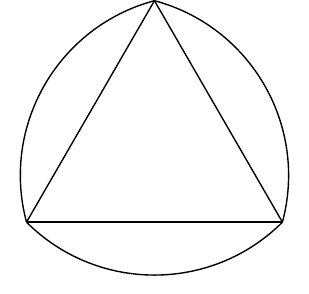}+
\includegraphics[scale=0.6,align=c]{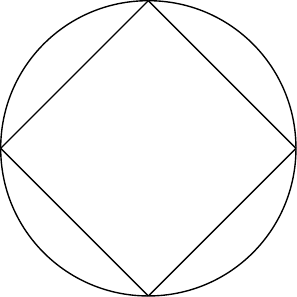}+
\includegraphics[scale=0.47,align=c]{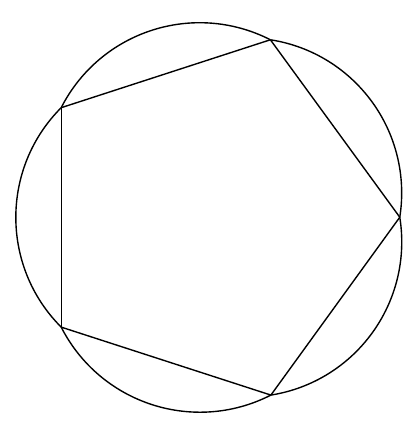}+\,\dots
\]
In the high-temperature limit the Mandelstam variables are constrained like
\[
|s|,|t|,|u|\leq|\phi|^2=\frac{\Lambda^3}{2\pi T}\propto 1/T\rightarrow0 \,.
\]
Hence, for $\lambda\gg\lambda_c$ it is then sufficient to consider $f(\lambda,0)\equiv f(\lambda)$ which is independent of the loop integrations and can be factored out in the DS equation, namely
\begin{align}
&\includegraphics[scale=0.8,align=c]{feynman-graphs/DS-closed-1} 
=\includegraphics[scale=0.8,align=c]{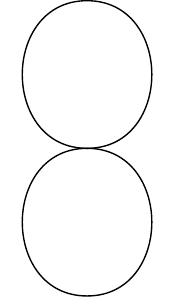} 
+\includegraphics[scale=0.8,align=c]{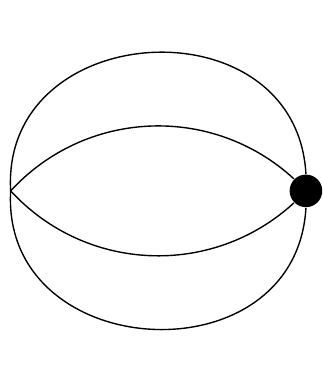}\\
\Rightarrow\qquad&
f(\lambda)\Delta P|_\mathrm{2-loop} = \Delta P|_\mathrm{2-loop} + f(\lambda)\Delta P|_\mathrm{3-loop}\,.
\end{align}
Solving for $f(\lambda)$ yields
\[\label{eq:foflambda}
f(\lambda) = \frac{\Delta P|_\mathrm{2-loop}}{\Delta P|_\mathrm{2-loop} - \Delta P|_\mathrm{3-loop}}\approx  -\num{0.94E15}i\lambda^{-11.6}\,,
\]
where in the final step we worked to leading order in $\lambda$ and used for $\Delta P|_\mathrm{2-loop}$ a fit to numerical data between $\lambda=200=14.42\lambda_c$ and $\lambda=1000=72.10\lambda_c$ which yields  $\Delta P|_\mathrm{2-loop}=\num{-5E-5}\Lambda^4\lambda^{1.4}$ and for $\Delta P|_\mathrm{3-loop}$ we used the power law of \eqref{eq:approxintss3}.
The fact that $\Delta P|_\mathrm{2-loop}$ is real while $\Delta P|_\mathrm{3-loop}$ is imaginary ensures that $f(\lambda)$ is free of singularities. Using this result to calculate the two-loop and three-loop contributions with resummed vertices now yields the well-bounded results
\[\label{eq:3lresummed}
\begin{split}
f^2(\lambda)\Delta P|_\mathrm{3-loop}&=-\num{4.7E10}i\Lambda^4\lambda^{-10.2}\\
f(\lambda)\Delta P|_\mathrm{2-loop}&=\num{4.7E10}i\Lambda^4\lambda^{-10.2}
\end{split}
\]
to leading order in $\lambda$, implying that these leading orders exactly cancel. Subleading order contributions are thus safely bounded and contain imaginary contributions. We interpret the small imaginary contributions as non-thermal modifications of the thermodynamically self-consistent one-loop pressure. Their origin may be inhomogeneities in the thermal ground state and thus the packing voids between densely packed (anti)caloron centers.
A rather reassuring observation is that if one postulates that the fractional form of $f(\lambda)$ in \eqref{eq:foflambda} persists down to low temperatures, this would imply that $f(\lambda)\approx1$ close to $\lambda_c$ which would be consistent with the hierarchy displayed already at the non-resummed level as illustrated in \autoref{fig:3l-vs-2l}. 

\section{Summary and conclusions}\label{sec:4}
We aimed in this work to provide an insight into how radiative corrections beyond two-loop order to the thermal ground state of SU(2) Yang-Mills theory can be organised. The vertex constraints arising from the thermal ground state have been demonstrated to be insufficient to reduce the loop expansion to a finite number of diagrams. Moreover, explicit calculation of the 2PI three-loop diagram in the massive sector showed that these constraints are also not strong enough to extend the hierarchy in loop orders observed at low temperatures up to high temperatures. 
Resummation of corresponding classes of diagrams, however, has been demonstrated to be a promising resolution to this problem, yielding well-bounded corrections at all temperatures. The arising small non-thermal (imaginary) corrections to the pressure have been interpreted as a result of inhomogenities in the thermal ground state constituted of densely packed centers of Harrington-Shepard (anti)calorons.
At this stage it is not yet clear if further 2PI bubble diagrams in the massive sector are sufficiently constrained prior to resummation, due to lower symmetry and hence likely lower number of possibly equivalent constraints (cf. \autoref{sec:sign-constraints}), or if more resummation procedures are necessary and possible to control the expansion. For an exhaustive understanding of the radiative corrections, the massless and mixed sectors will also have to be treated in a similar manner.

The subject of how to organise the computation of radiative corrections in deconfining Yang-Mills thermodynamics thus is a broad one. Being of immediate urgency, it would be important to analyse diagrams symmetric under the $n$-th dihedral group (cf. \autoref{sec:sign-constraints}) that are generated by one massless and one massive propagator per bubble.

\section*{Acknowledgments}
I thank my collaborators Thierry Grandou and Ralf Hofmann for their contributions to the related paper \cite{BGH}. Furthermore I thank ITP of Heidelberg University for the funding of a two-week stay at INLN (Nice) in September 2016 during which, among other projects, this work was pursued and both ITP Heidelberg and INPHYNI (Nice) for funding my participation at the 5th winter workshop on non perturbative QFT (2017) at INPHYNI, where these results were presented.

The author declares that there is no conflict of interest regarding the publication of this paper. This includes the above mentioned funding.

\bibliographystyle{hunsrt}
\bibliography{thesisbib}{}

\end{document}